\documentclass[12pt,letterpaper]{article}
\usepackage{amsmath,amssymb,array,calc,rotating,epsfig,psfrag,amscd, cite}

\numberwithin{equation}{section}

\makeatletter
\renewcommand\section{\@startsection {section}{1}{\z@}%
                                   {-3.5ex \@plus -1ex \@minus -.2ex}
                                   {2.3ex \@plus.2ex}%
                                   {\normalfont\large\bfseries}}
\renewcommand\subsection{\@startsection{subsection}{2}{\z@}%
                                     {-3.25ex\@plus -1ex \@minus -.2ex}%
                                     {1.5ex \@plus .2ex}%
                                     {\normalfont\bfseries}}
\makeatother

\let\non\nonumber

\let\s=\sigma

\let\S=\Sigma

\newcommand{\bea}{\begin{eqnarray}}
\newcommand{\eea}{\end{eqnarray}}
\newcommand{\be}{\begin{equation}}
\newcommand{\ee}{\end{equation}}

\newcommand{\m}{\mu}

\newcommand{\p}{\partial}

\newcommand{\C}[1]{$(\ref{#1})$}

\def\IZ{\relax\ifmmode\mathchoice
{\hbox{\cmss Z\kern-.4em Z}}{\hbox{\cmss Z\kern-.4em Z}}
{\lower.9pt\hbox{\cmsss Z\kern-.4em Z}} {\lower1.2pt\hbox{\cmsss
Z\kern-.4em Z}}\else{\cmss Z\kern-.4em Z}\fi}
\def\IR{\relax{\rm I\kern-.18em R}}

\def\one{{\hbox{ 1\kern-.8mm l}}}

\newlength{\bredde}
\def\slash#1{\settowidth{\bredde}{$#1$}\ifmmode\,\raisebox{.15ex}{/}
\hspace*{-\bredde} #1\else$\,\raisebox{.15ex}{/}\hspace*{-\bredde}
#1$\fi}

\newsavebox{\zzzbar}
\sbox{\zzzbar}
  {\setlength{\unitlength}{0.9em}
  \begin{picture}(0.6,0.7)
  \thinlines
  \put(0,0){\line(1,0){0.6}}
  \put(0,0.75){\line(1,0){0.575}}
  \multiput(0,0)(0.0125,0.025){30}{\rule{0.3pt}{0.3pt}}
  \multiput(0.2,0)(0.0125,0.025){30}{\rule{0.3pt}{0.3pt}}
  \put(0,0.75){\line(0,-1){0.15}}
  \put(0.015,0.75){\line(0,-1){0.1}}
  \put(0.03,0.75){\line(0,-1){0.075}}
  \put(0.045,0.75){\line(0,-1){0.05}}
  \put(0.05,0.75){\line(0,-1){0.025}}
  \put(0.6,0){\line(0,1){0.15}}
  \put(0.585,0){\line(0,1){0.1}}
  \put(0.57,0){\line(0,1){0.075}}
  \put(0.555,0){\line(0,1){0.05}}
  \put(0.55,0){\line(0,1){0.025}}
  \end{picture}}

\newcommand{\ena}{\end{eqnarray}}
\newcommand{\beqa}{\begin{eqnarray}}
\newcommand{\eeqa}{\end{eqnarray}}

\def\m{\mu}

\def\s{\sigma}

\def\S{\Sigma}

\begin{document}
\begin{titlepage}

\begin{center}



\vskip 2 cm
{\Large \bf Integrating simple genus two string invariants over moduli space}\\
\vskip 1.25 cm { Anirban Basu\footnote{email address:
    anirbanbasu@hri.res.in} } \\
{\vskip 0.5cm  Harish--Chandra Research Institute, HBNI, Chhatnag Road, Jhusi,\\
Prayagraj 211019, India}

\end{center}

\vskip 2 cm

\begin{abstract}
\baselineskip=18pt

We consider an $Sp(4,\mathbb{Z})$ invariant expression involving two factors of the Kawazumi--Zhang (KZ) invariant each of which is a modular graph with one link, and four derivatives on the moduli space of genus two Riemann surfaces. Manipulating it, we show that the integral over moduli space of a linear combination of a modular graph with two links and the square of the KZ invariant reduces to a boundary integral. We also consider an $Sp(4,\mathbb{Z})$ invariant expression involving three factors of the KZ invariant and six derivatives on moduli space, from which we deduce that the integral over moduli space of a modular graph with three links reduces to a boundary integral. In both cases, the boundary term is completely determined by the KZ invariant. We show that both the integrals vanish.

\end{abstract}

\end{titlepage}


\section{Introduction}

Multiloop scattering amplitudes involving massless external states in perturbative string theory contain very useful information about the structure of the effective action. While in general a detailed analysis of such amplitudes is difficult to perform, the analysis simplifies in compactifications which preserve large amount of supersymmetry. The structure of the amplitudes involving gravitons as external states in toroidally compactified type II string theory, which preserves maximal supersymmetry, has been analyzed at genus one as well as at genus two, while very little is known beyond. The $\alpha'$ expansion of these amplitudes yields terms that are analytic as well as non--analytic in the external momenta in the effective action. In order to obtain the precise coefficients of the various analytic terms, one has to integrate over the entire moduli space of the Riemann surface, where the integrand is built out of string invariants or modular graph forms~\cite{DHoker:2015gmr,DHoker:2015wxz}. These graphs have links given by the scalar Green function or its worldsheet derivative, while the vertices are the positions of insertions of the vertex operators on the worldsheet. Hence analyzing various properties of these string invariants plays a crucial role in determining terms in the effective action. This has led to a detailed analysis of their properties~\cite{Green:1999pv,DHoker:2005vch,DHoker:2005jhf,Berkovits:2005df,Berkovits:2005ng,Green:2008uj,Richards:2008jg,Green:2013bza,DHoker:2013fcx,DHoker:2014oxd,Pioline:2015qha,DHoker:2015gmr,DHoker:2015wxz,Basu:2015ayg,DHoker:2016mwo,Basu:2016xrt,Basu:2016kli,Basu:2016mmk,DHoker:2016quv,Kleinschmidt:2017ege,DHoker:2017pvk,DHoker:2018mys,DHoker:2019blr,Basu:2019idd,Gerken:2019cxz,Gerken:2020yii,DHoker:2020prr,Gerken:2020aju,DHoker:2020tcq,Basu:2020pey,DHoker:2020uid,Basu:2020iok,Gerken:2020xfv,DHoker:2020aex} revealing a rich underlying structure.           

In this paper, rather than directly analyze the string invariants, we shall be concerned with performing their integrals over moduli space at genus two.  Such integrals have been considered at genus one and two leading to various interactions in the effective action~\cite{Green:1999pv,DHoker:2005vch,DHoker:2005jhf,Green:2008uj,Richards:2008jg,Green:2013bza,DHoker:2014oxd,DHoker:2015gmr,Basu:2015dqa,Basu:2016fpd,DHoker:2019blr,DHoker:2020tcq}. While at genus one, these integrals are over the fundamental domain of $SL(2,\mathbb{Z})$, at genus two, they are over the fundamental domain of $Sp(4,\mathbb{Z})$. 

Let us consider the terms in the effective action that arise in the low momentum expansion of the four graviton amplitude at genus two~\cite{DHoker:2005vch}. The leading contribution which is  the $D^4\mathcal{R}^4$ interaction involves simply integrating over the volume element of the fundamental domain of $Sp(4,\mathbb{Z})$~\cite{DHoker:2005jhf}. At the next order in the $\alpha'$ expansion, we have the $D^6\mathcal{R}^4$ interaction which involves integrating the Kawazumi--Zhang (KZ) invariant over moduli space~\cite{Kawazumi,Zhang,DHoker:2013fcx}, which is done by reducing it to a boundary term using an eigenvalue equation the KZ invariant satisfies~\cite{DHoker:2014oxd}. While the structure of the string invariants that arise in the integrand of the $D^8\mathcal{R}^4$ interaction has been analyzed in asymptotic expansions around the degenerating nodes~\cite{DHoker:2017pvk,DHoker:2018mys}, their integrals over moduli space have not been performed. Similar is the status of interactions that arise in the low momentum expansion of the five graviton amplitude~\cite{DHoker:2020prr,DHoker:2020tcq}. While the KZ invariant yields a graph with only one link, the integrands for the amplitudes at higher orders in the $\alpha'$ expansion involve a sum of terms each of which has graphs with a total of at least two links, where the number of links in such terms increases as one goes to higher and higher orders in the $\alpha'$ expansion.    

Thus it is important to understand how to perform the integrals over moduli space involving integrands given by string invariants for interactions that are $\alpha'$ suppressed compared to the $D^6\mathcal{R}^4$ interaction in the low momentum expansion of the four graviton amplitude. The graphs that arise in this expansion to all orders in the $\alpha'$ expansion have links that are given by the Green function. The situation gets more involved when one considers the graphs that arise in the low momentum expansion of the five graviton amplitude, where additional contributions arise involving graphs with links given by the worldsheet derivative of the Green function. Thus it is interesting in general to understand the issue of integrating various string invariants over moduli space.  

One of the major obstacles in performing these integrals over moduli space at genus two arises from the fact that beyond the graph for the $D^6\mathcal{R}^4$ interaction, there are no known eigenvalue equations involving the Laplacian operator on moduli space the string invariants satisfy that are useful in performing the integrals\footnote{The eigenvalue equation obtained in~\cite{Basu:2018bde} is trivially satisfied on using the identities derived in~\cite{DHoker:2020tcq}. I am thankful to Boris Pioline for useful comments on this issue.}.    

In this paper, we shall perform the integrals over moduli space for certain simple genus two graphs. They are simple in the sense that the links in them are disconnected\footnote{This does not mean that such graphs always factorize in terms of graphs with lesser number of links which follows from their detailed structure.} and hence they do not form closed loops on the worldsheet. We first consider the integral of a graph with two links that arises in the analysis of the $D^8\mathcal{R}^4$ term in the low momentum expansion. We show that the integral with the $Sp(4,\mathbb{Z})$ invariant measure of a linear combination of this graph and the square of the KZ invariant reduces to a boundary term on moduli space. Hence the integral can be evaluated based on only the knowledge of the asymptotic expansions around the separating and non--separating nodes of the boundary contribution, which turns out to be completely determined by the KZ invariant. We next perform a similar analysis for a simple graph with three links that should arise in the analysis of the $D^6\mathcal{R}^6$ term in the low momentum expansion of the six graviton amplitude. We evaluate the integral by reducing it to a boundary contribution, which again is entirely determined by the KZ invariant. In both cases, the integral vanishes.

For the analysis of the graphs with two links, we start with an $Sp(4,\mathbb{Z})$ invariant expression involving two factors of the KZ invariant and four derivatives on moduli space. Using the differential equation satisfied by the KZ invariant, this expression reduces to a graph with two links and no worldsheet derivatives. Proceeding differently, we next manipulate this expression to reduce it to a boundary term along with an additional contribution involving the square of the KZ invariant. Equating the two results obtained by evaluating the same expression differently yields our desired answer. In obtaining these relations which are valid everywhere in the bulk of moduli space, we use the eigenvalue equation the KZ invariant satisfies, as well as the identities deduced in~\cite{DHoker:2020tcq}. We next perform a similar analysis using an $Sp(4,\mathbb{Z})$ invariant expression involving three factors of the KZ invariant and six derivatives on moduli space, leading to our desired answer. The intermediate steps involve obtaining several algebraic relations between simple graphs with three links which we obtain separately. We expect the analysis to generalize to simple graphs with arbitrary number of links.               

We begin by reviewing facts about genus two string amplitudes that are relevant for our purposes. We then perform the analysis everywhere in the bulk of moduli space for graphs with two links, and then for graphs with three links. Finally, we perform the integrals over moduli space by evaluating the boundary contributions.

\section{Genus two string amplitudes and the Kawazumi--Zhang invariant}

We denote the genus two worldsheet by $\S_2$, and the conformally invariant Arakelov Green function by $G(z,w)$. The period matrix is defined by $\Omega_{IJ} = X_{IJ} + iY_{IJ}$ ($I,J=1,2$), where the matrices $X$ and $Y$ have real entries. Also we define $Y^{-1}_{IJ} = (Y^{-1})_{IJ}$, as well as the dressing factors
\be (z,\overline{w}) = Y^{-1}_{IJ} \omega_I (z) \overline{\omega_J (w)}, \quad \mu (z) = (z,\overline{z}), \quad P(z,w) = (z,\overline{w})(w,\overline{z}),\ee
where $\omega_I = \omega_I (z) dz$ is the Abelian differential one form. Every string invariant is given by a graph with links involving the Arakelov Green function, along with a specific choice of dressing factors for the integrated vertices.  
The integration measure over the worldsheet is given by $d^2 z = i dz \wedge d\overline{z} = 2 d({\rm Re}z)\wedge d({\rm Im}z)$. 

The Kawazumi--Zhang invariant which appears in the analysis of the $D^6\mathcal{R}^4$ interaction is given by
\be \label{KZ}\mathcal{B}_1 (\Omega,\overline\Omega) = \int_{\S_2^2} \prod_{i=1}^2 d^2 z_i G(z_1,z_2) P(z_1,z_2)\ee
as depicted by figure 1\footnote{In the various figures depicting the graphs, the solid and dashed lines represent the Green function and the dressing factor connecting the vertices respectively.}. We now write down several expressions involving it that are satisfied everywhere in the bulk of moduli space, which are very useful for our purposes. 
  
\begin{figure}[ht]
\begin{center}
\[
\mbox{\begin{picture}(130,45)(0,0)
\includegraphics[scale=.6]{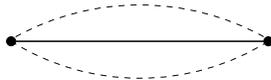}
\end{picture}}
\]
\caption{The string invariant $\mathcal{B}_1$}
\end{center}
\end{figure}

Defining 
\be \label{pdef}\p_{IJ} = \frac{1}{2} \Big(1+\delta_{IJ}\Big)\frac{\p}{\p\Omega_{IJ}}\ee 
we see that
\be \label{bulkrel}\p_{KL}\Omega_{IJ} = \frac{1}{2}\Big(\delta_{IK}\delta_{JL} + \delta_{IL}\delta_{JK}\Big)\ee
everywhere in the bulk of moduli space. 
The KZ invariant \C{KZ} satisfies the differential equation~\cite{DHoker:2014oxd}
\be \label{main}\p_{IJ} \overline\p_{KL} \mathcal{B}_1 = \frac{1}{16}\Big(\Theta_{IJ;KL} + \Theta_{IJ;LK} + \Theta_{JI;KL} + \Theta_{JI;LK}\Big),\ee
where $\Theta_{IJ;KL}$ is defined by
\bea \label{Theta}\Theta_{IJ;KL} &=& 5 Y^{-1}_{AK} Y^{-1}_{IB} \int_{\S_2^2} \prod_{i=1}^2 d^2 z_i G(z_1,z_2) \omega_A (z_1) \overline{\omega_B (z_2)} \non \\ && \times \Big(Y^{-1}_{JL}(z_2,\overline{z}_1)- Y^{-1}_{LC} \omega_C (z_2) Y^{-1}_{JD} \overline{\omega_D (z_1)}\Big).  \eea
We shall see that \C{main} will play a central role in our analysis.   

Defining the $Sp(4,\mathbb{Z})$ invariant Laplacian by
\be \label{L}\Delta = 4Y_{IK} Y_{JL}\p_{IJ}\overline\p_{KL}, \ee
we find that \C{KZ} satisfies the eigenvalue equation
\be \label{eigenKZ}\Big(\Delta -5\Big) \mathcal{B}_1 =0\ee
in the bulk of moduli space, which we shall often use. In obtaining \C{eigenKZ}, we have used the vanishing integral
\be \label{zero}\int_{\S} d^2 z \m(z) G(z,w)=0\ee
the Arakelov Green function satisfies.

Now the Siegel upper half space $\mathcal{H}_2$, where the various amplitudes are naturally defined,  is K$\ddot{\rm{a}}$hler, and the $Sp(4,\mathbb{R})$ invariant K$\ddot{\rm{a}}$hler metric is
\be ds^2 = Y^{-1}_{IJ} Y^{-1}_{KL} d\Omega_{IK} d\overline\Omega_{JL}.\ee
Thus from the inverse metric we see that $Y_{IJ}$ naturally contracts with one holomorphic and one anti--holomorphic index in moduli space. Let us try to construct $Sp(4,\mathbb{Z})$ invariants while insisting that we only allow such contractions.

Hence using only $\p\overline\p \mathcal{B}_1$ involving derivatives over moduli space to construct an $Sp(4,\mathbb{Z})$ invariant, we end up with $\Delta \mathcal{B}_1$ using the definition \C{L}, which is the only possibility.

\section{Analyzing simple string invariants with two links}

Based on the discussion above, let us try to construct invariants of the form $(\p\overline\p \mathcal{B}_1)^2$. In fact, there are two possibilities for constructing invariants that do not factorize. We define one of them by\footnote{The other one is given by
\be \label{drop}Y_{AC}Y_{BK}Y_{DI}Y_{JL}\Big(\p_{AB}\overline\p_{CD}\mathcal{B}_1\Big)\Big(\p_{IJ}\overline\p_{KL}\mathcal{B}_1\Big)\ee
which we do not consider.}
\be \label{defchi}\chi^{(2)} (\Omega,\overline\Omega)= 16 Y_{AK} Y_{BL} Y_{CI} Y_{DJ}\Big(\p_{AB} \overline\p_{CD} \mathcal{B}_1 \Big)\Big(\p_{IJ} \overline\p_{KL} \mathcal{B}_1\Big).\ee
Thus preserving the symmetries, from \C{main} we have that
\bea \chi^{(2)} = \frac{1}{4}\Big(Y_{AK}Y_{BL}+ Y_{AL} Y_{BK}\Big) \Big( Y_{CI}Y_{DJ}+ Y_{CJ}Y_{DI}\Big) \Theta_{AB;CD} \Theta_{IJ;KL}.\eea

Using the expression \C{Theta}, this gives us that
\bea \label{eval}\chi^{(2)} = \frac{25}{4}\Big(2 \mathcal{B}_2 + 2 \mathcal{B}_3 + \mathcal{B}_1^2\Big),\eea
on using \C{zero}. In \C{eval}, the two string invariants are given by
\bea \mathcal{B}_2 &=& \int_{\S^4} \prod_{i=1}^4 d^2 z_i G(z_1,z_2) G(z_3,z_4) P(z_1,z_3)P(z_2,z_4), \non \\ \mathcal{B}_3 &=&  \int_{\S^4} \prod_{i=1}^4 d^2 z_i G(z_1,z_2) G(z_3,z_4) (z_1,\overline{z_4})(z_4,\overline{z_2})(z_2,\overline{z_3})(z_3,\overline{z_1}). \eea
In the intermediate steps of the analysis, we also come across the string invariant
\bea \label{defB4}\mathcal{B}_4 =  \int_{\S^4} \prod_{i=1}^4 d^2 z_i G(z_1,z_2) G(z_3,z_4) (z_1,\overline{z_3})(z_3,\overline{z_4})(z_4,\overline{z_2})(z_2,\overline{z_1})\eea
which cancels in the final answer. While the graph $\mathcal{B}_2$ arises in the low momentum expansion of the four and five graviton amplitudes, the graphs $\mathcal{B}_3$ and $\mathcal{B}_4$ arise in the low momentum expansion of the five graviton amplitude. These three graphs, depicted by figure 2, differ only in their dressing factors. 

\begin{figure}[ht]
\begin{center}
\[
\mbox{\begin{picture}(270,130)(0,0)
\includegraphics[scale=.65]{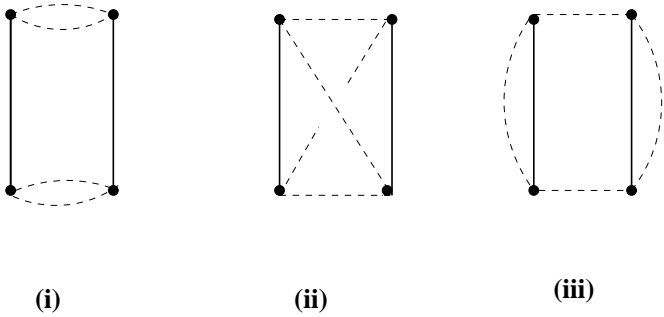}
\end{picture}}
\]
\caption{The string invariants (i) $\mathcal{B}_2$, (ii) $\mathcal{B}_3$ and (iii) $\mathcal{B}_4$}
\end{center}
\end{figure}

Using the relation between the graphs $\mathcal{B}_2$ and $\mathcal{B}_3$ given by~\cite{DHoker:2020tcq}
\be \label{rel23}\mathcal{B}_3 = \mathcal{B}_2 - \frac{\mathcal{B}_1^2}{2},\ee
from \C{eval} we get that
\be \label{chirel}\chi^{(2)} =  25 \mathcal{B}_2\ee
yielding a simple string invariant with two links. Thus we obtain this result by evaluating $\chi^{(2)}$ directly.

Let us now evaluate $\chi^{(2)}$ differently. First manipulating $\overline\p_{CD}$ in \C{defchi}, we express $\chi^{(2)}$ as\footnote{In obtaining this as well as similar results below, we use \C{bulkrel} heavily.}
\bea \label{1}\frac{\chi^{(2)}}{16} &=& ({\rm det}Y)^3 \overline\p_{CD}\Big[({\rm det}Y)^{-3}Y_{AK}Y_{BL}Y_{CI}Y_{DJ}\Big(\p_{AB}\mathcal{B}_1\Big)\Big(\p_{IJ}\overline\p_{KL}\mathcal{B}_1\Big)\Big]\non \\ &&-\frac{Y_{IK}Y_{JL}}{4} \Big(\p_{IJ} \mathcal{B}_1 \Big)\Big(\overline\p_{KL} \Delta \mathcal{B}_1\Big).\eea
In fact using \C{eigenKZ}, the last term in \C{1} is equal to
\be \label{rem}-\frac{5Y_{IK}Y_{JL}}{4} \Big(\p_{IJ} \mathcal{B}_1 \Big)\Big(\overline\p_{KL} \mathcal{B}_1\Big).\ee
Now in \C{rem}, we further manipulate $\overline\p_{KL}$ to obtain
\bea \label{2}Y_{IK}Y_{JL} \Big(\p_{IJ} \mathcal{B}_1 \Big)\Big(\overline\p_{KL} \mathcal{B}_1\Big) = ({\rm det}Y)^3 \overline\p_{KL} \Big[ ({\rm det}Y)^{-3} Y_{IK} Y_{JL} \mathcal{B}_1\Big(\p_{IJ}\mathcal{B}_1\Big)\Big] - \frac{1}{4} \mathcal{B}_1 \Delta \mathcal{B}_1.\eea
The last term in \C{2} is equal to $-5\mathcal{B}_1^2/4$ on using \C{eigenKZ}. Hence putting the various contributions together, we obtain an alternate expression for $\chi^{(2)}$. Equating this expression with \C{chirel}, we get that
\bea \label{simplify1}&&\frac{25}{16} \Big(\mathcal{B}_2 - \mathcal{B}_1^2\Big) = -\frac{5}{4}({\rm det}Y)^3 \overline\p_{KL} \Big[ ({\rm det} Y)^{-3}Y_{IK}Y_{JL} \mathcal{B}_1 \Big(\p_{IJ} \mathcal{B}_1\Big)\Big]\non \\ &&+ ({\rm det}Y)^3\overline\p_{CD} \Big[({\rm det}Y)^{-3}Y_{AK} Y_{BL} Y_{CI} Y_{DJ} \Big(\p_{AB} \mathcal{B}_1\Big)\Big(\p_{IJ} \overline\p_{KL}\mathcal{B}_1\Big) \Big].\eea
Thus we see that $\mathcal{B}_2$ is entirely determined by the KZ invariant $\mathcal{B}_1$.
In fact the combination $(\mathcal{B}_2 - \mathcal{B}_1^2)/{({\rm det}Y)}^3$ is a total derivative in moduli space, which will be very helpful for us later. 

We can simplify \C{simplify1} by using 
\be \label{YY}Y_{IJ} Y_{KL} - Y_{IL}Y_{JK} = \epsilon_{IK} \epsilon_{JL}({\rm det}Y)\ee
and \C{eigenKZ}.
This leads to
\bea  \label{g1}\frac{25\Big(\mathcal{B}_2 - \mathcal{B}_1^2\Big)}{16({\rm det}Y)^3}  &=& \epsilon_{AI}\epsilon_{BJ}\epsilon_{CK} \epsilon_{DL}\overline\p_{CD} \Big[ \frac{\Big(\p_{AB} \mathcal{B}_1\Big) \Big(\p_{IJ} \overline\p_{KL}\mathcal{B}_1\Big)}{{\rm det}Y}\Big]\non \\ &&- 2\epsilon_{AI}\epsilon_{CK}\overline\p_{CD} \Big[\frac{Y_{BD}  Y_{JL}}{({\rm det}Y)^2} \Big(\p_{AB} \mathcal{B}_1\Big)\Big(\p_{IJ} \overline\p_{KL}\mathcal{B}_1\Big) \Big].\eea

It will be interesting to analyze the invariant \C{drop} by proceeding along the same lines to see what we obtain, though the details are going to be more involved. To see this, note that in the above analysis involving $\chi^{(2)}$, very schematically manipulating the derivatives we have obtained
\be \chi^{(2)} \sim \Big(Y_{AK} Y_{BL}\p_{AB} \overline\p_{KL}\mathcal{B}_1\Big)\Big( Y_{CI}Y_{DJ}\p_{IJ} \overline\p_{CD} \mathcal{B}_1\Big)+\ldots \sim \frac{(\Delta \mathcal{B}_1)^2}{16} +\ldots \sim \frac{25 \mathcal{B}_1^2}{16}+\ldots,\ee     
leading to the final expression, where the terms we have ignored involve total derivatives on moduli space up to an overall factor of $({\rm det}Y)^3$. Such a simplification does not occur in the analysis of \C{drop} given the index structure of the invariant.    

\section{Analyzing simple string invariants with three links}

Let us now perform a similar analysis involving graphs with three links, where we construct invariants of the type $(\p\overline\p \mathcal{B}_1)^3$. While there are four such invariants one can consider that do not factorize, we only focus on the one we define by\footnote{The other three invariants are given by
\bea \label{drop2}Y_{AK}Y_{BL}Y_{CI}Y_{JP}Y_{DM}Y_{NQ}\Big(\p_{AB} \overline\p_{CD} \mathcal{B}_1 \Big)\Big(\p_{IJ} \overline\p_{KL} \mathcal{B}_1\Big)\Big(\p_{MN} \overline\p_{PQ} \mathcal{B}_1\Big) , \non \\ Y_{AC}Y_{BK}Y_{DI}Y_{JP}Y_{LM}Y_{NQ}\Big(\p_{AB} \overline\p_{CD} \mathcal{B}_1 \Big)\Big(\p_{IJ} \overline\p_{KL} \mathcal{B}_1\Big)\Big(\p_{MN} \overline\p_{PQ} \mathcal{B}_1\Big), \non \\ Y_{AC}Y_{BK}Y_{IL}Y_{JP}Y_{DM}Y_{NQ}\Big(\p_{AB} \overline\p_{CD} \mathcal{B}_1 \Big)\Big(\p_{IJ} \overline\p_{KL} \mathcal{B}_1\Big)\Big(\p_{MN} \overline\p_{PQ} \mathcal{B}_1\Big).\eea
Along the lines of the discussion in the previous section, we see that the analysis of these invariants is going to be more involved than the analysis of \C{firstdef}. As we shall soon analyze in detail, manipulating the derivatives in $\chi^{(3)}$ yield terms of the form $\mathcal{B}_1^3$ and $\mathcal{B}_1\mathcal{B}_2$ as well as other contributions that yield total derivatives on moduli space, up to an overall factor of $({\rm det}Y)^3$. This simplification does not happen for the invariants in \C{drop2} given their index structure.

In fact, the natural generalization of $\chi^{(2)}$ and $\chi^{(3)}$ involving $n$ factors each of $\mathcal{B}_1$, $\p_{AB}$ and $\overline\p_{CD}$ is given by
\bea \label{chin}\chi^{(n)} (\Omega,\overline\Omega) = 4^n \prod_{i=1}^n \Big(Y_{A_i C_{i+1}} Y_{B_i D_{i+1}}\Big)\prod_{j=1}^n \Big(\p_{A_j B_j} \overline\p_{C_j D_j}\mathcal{B}_1 \Big),\eea
where $C_{n+1} \equiv C_1$ and $D_{n+1} \equiv D_1$. Among the several string invariants that arise at this order, we expect the manipulations involving \C{chin} to be the simplest given the index structure.}
\bea \label{firstdef}\chi^{(3)} (\Omega,\overline\Omega) = 64 Y_{AK}Y_{BL}Y_{IP}Y_{JQ}Y_{CM}Y_{DN}\Big(\p_{AB} \overline\p_{CD} \mathcal{B}_1 \Big)\Big(\p_{IJ} \overline\p_{KL} \mathcal{B}_1\Big)\Big(\p_{MN} \overline\p_{PQ} \mathcal{B}_1\Big).\eea

We first calculate $\chi^{(3)}$ using \C{main}. Keeping the symmetries manifest, we get that
\bea \chi^{(3)} &=& \frac{1}{8} \Big(Y_{AK}Y_{BL}+Y_{AL}Y_{BK}\Big)\Big(Y_{IP}Y_{JQ}+Y_{IQ}Y_{JP}\Big)\Big(Y_{CM}Y_{DN}+Y_{CN}Y_{DM}\Big)\non \\ &&\times \Theta_{AB;CD} \Theta_{IJ;KL} \Theta_{MN;PQ}.\eea
Then using \C{Theta} we obtain
\bea \label{finval}\chi^{(3)} = \frac{125}{8} \Big[12 \Big(\mathcal{B}_5 +  \mathcal{B}_6 \Big)- 6 \Big(\mathcal{B}_7 + \mathcal{B}_8 \Big)-4\Big( \mathcal{B}_9 + \mathcal{B}_{10}\Big) + 3\mathcal{B}_1 \mathcal{B}_4 \Big],\eea
where the various string invariants are given by\footnote{We expect the $\mathcal{R}^6$ and $D^2\mathcal{R}^6$ interactions to be related to the 1/4 BPS $D^4\mathcal{R}^4$ and 1/8 BPS $D^6\mathcal{R}^4$ interactions respectively, and hence have the same string invariants as their integrands. However, we expect that the integrands for the non--BPS $D^4\mathcal{R}^6$ and $D^6\mathcal{R}^6$ interactions should contain additional graphs beyond those that arise in the integrands of the non--BPS $D^8\mathcal{R}^4$ and $D^{10}\mathcal{R}^4$ interactions respectively. Thus we expect that the $D^6\mathcal{R}^6$ amplitude should contain in its integrand disconnected graphs with three links involving six vertices. Hence we refer to them as string invariants.}
\bea \label{defB}\mathcal{B}_5 &=& \int_{\S^6} \prod_{i=1}^6 d^2 z_i G(z_1,z_2)G(z_3,z_4)G(z_5,z_6)(z_5,\overline{z_6})(z_6,\overline{z_3})(z_3,\overline{z_1})(z_1,\overline{z_5})P(z_2,\overline{z_4}), \non \\  \mathcal{B}_6 &=&  \int_{\S^6} \prod_{i=1}^6 d^2 z_i G(z_1,z_2)G(z_3,z_4)G(z_5,z_6)(z_5,\overline{z_6})(z_6,\overline{z_3})(z_3,\overline{z_2})(z_2,\overline{z_4})(z_4,\overline{z_1})(z_1,\overline{z_5}), \non \\  \mathcal{B}_7 &=& \int_{\S^6} \prod_{i=1}^6 d^2 z_i G(z_1,z_2)G(z_3,z_4)G(z_5,z_6)(z_1,\overline{z_3})(z_3,\overline{z_5})(z_5,\overline{z_6})(z_6,\overline{z_4})(z_4,\overline{z_2})(z_2,\overline{z_1}),\non \\ \mathcal{B}_8 &=&  \int_{\S^6} \prod_{i=1}^6 d^2 z_i G(z_1,z_2)G(z_3,z_4)G(z_5,z_6)(z_2,\overline{z_1})(z_1,\overline{z_3})(z_3,\overline{z_2})(z_6,\overline{z_5})(z_5,\overline{z_4})(z_4,\overline{z_6}),\non \\  \mathcal{B}_9 &=&  \int_{\S^6} \prod_{i=1}^6 d^2 z_i G(z_1,z_2)G(z_3,z_4)G(z_5,z_6)(z_1,\overline{z_5})(z_5,\overline{z_3})(z_3,\overline{z_1})(z_2,\overline{z_6})(z_6,\overline{z_4})(z_4,\overline{z_2}), \non \\ \mathcal{B}_{10} &=&  \int_{\S^6} \prod_{i=1}^6 d^2 z_i G(z_1,z_2)G(z_3,z_4)G(z_5,z_6)(z_1,\overline{z_5})(z_5,\overline{z_3})(z_3,\overline{z_2})(z_2,\overline{z_6})(z_6,\overline{z_4})(z_4,\overline{z_1}).\non \\ \eea
In obtaining them, we have often used \C{zero}.

In fact the string invariant $\mathcal{B}_{11}$ defined by
\be \label{defB11}\mathcal{B}_{11} =  \int_{\S^6} \prod_{i=1}^6 d^2 z_i G(z_1,z_2)G(z_3,z_4)G(z_5,z_6)(z_5,\overline{z_6})(z_6,\overline{z_3})(z_3,\overline{z_4})(z_4,\overline{z_2})(z_2,\overline{z_1})(z_1,\overline{z_5})\ee
also arises in the intermediate stages of the calculation, but cancels in the final answer. Each of these simple string invariants involve three disconnected links, and only differ in their dressing factors. They are depicted by figure 3. 

\begin{figure}[ht]
\begin{center}
\[
\mbox{\begin{picture}(270,330)(0,0)
\includegraphics[scale=.7]{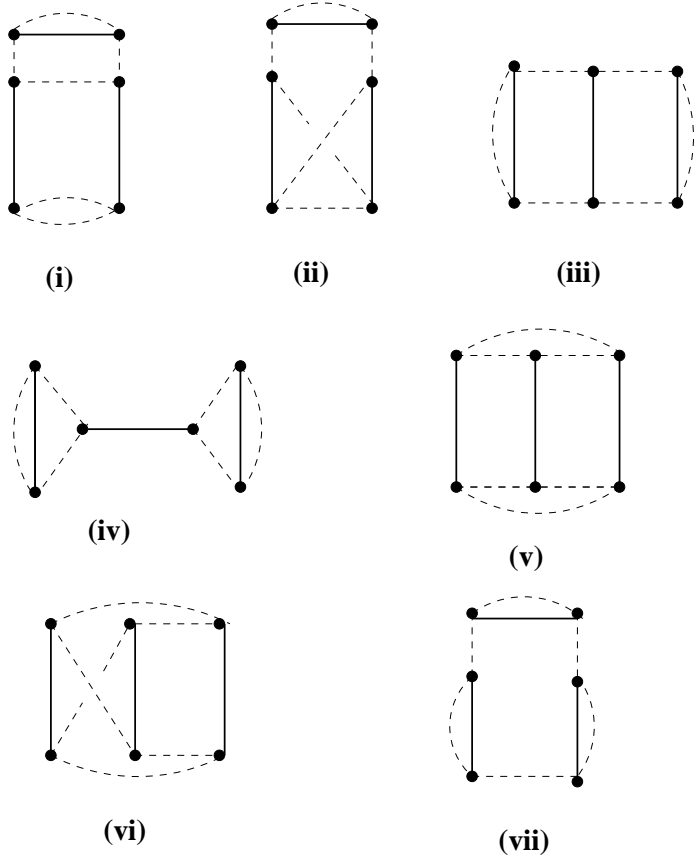}
\end{picture}}
\]
\caption{The string invariants (i) $\mathcal{B}_5$, (ii) $\mathcal{B}_6$, (iii) $\mathcal{B}_7$, (iv) $\mathcal{B}_8$, (v) $\mathcal{B}_9$, (vi) $\mathcal{B}_{10}$ and (vii) $\mathcal{B}_{11}$}
\end{center}
\end{figure}

To simplify the expression \C{finval}, we use the various relations between the graphs that have been deduced in appendix A, the relation \C{rel23} as well as the relation~\cite{DHoker:2020tcq}
\be \mathcal{B}_4= \frac{\mathcal{B}_1^2}{2}.\ee
Thus we see that \C{finval} yields
\bea\label{eq}\chi^{(3)} = \frac{125}{2} \Big[\mathcal{B}_1 \Big(3\mathcal{B}_2 -\mathcal{B}_1^2 \Big)- 2 \mathcal{B}_9\Big].\eea
Hence evaluating $\chi^{(3)}$ using \C{main}, we see that apart from graphs $\mathcal{B}_1$ and $\mathcal{B}_2$ that arise in the integrands of terms at lower orders in the $\alpha'$ expansion, it only depends on a single simple graph $\mathcal{B}_9$ with three links.

We now evaluate $\chi^{(3)}$ differently, where we often use \C{eigenKZ}. To start with, manipulating $\p_{AB}$ in \C{firstdef} and adding the complex conjugate contribution, we get that
\bea \label{3one} &&\frac{\chi^{(3)}}{32} = ({\rm det}Y)^3 \p_{AB} \Big[ ({\rm det}Y)^{-3} Y_{AK} Y_{BL}Y_{IP}Y_{JQ}Y_{CM}Y_{DN} \Big(\overline\p_{CD} \mathcal{B}_1\Big)\Big(\p_{IJ} \overline\p_{KL} \mathcal{B}_1\Big) \non \\ &&\times \Big(\p_{MN} \overline\p_{PQ}\mathcal{B}_1\Big)\Big] - \frac{5}{4}Y_{IP}Y_{JQ}Y_{CM}Y_{DN}\Big(\overline\p_{CD}\mathcal{B}_1\Big)\Big(\p_{IJ}\mathcal{B}_1\Big)\Big(\p_{MN}\overline\p_{PQ}\mathcal{B}_1\Big)\non \\&&-  Y_{AK}Y_{BL}Y_{IP}Y_{JQ}\Big(\overline\p_{CD} \mathcal{B}_1\Big)\Big(\p_{IJ}\overline\p_{KL}\mathcal{B}_1\Big)\p_{AB}\Big(Y_{CM}Y_{DN}\p_{MN}\overline\p_{PQ}\mathcal{B}_1\Big)+ c.c. .\eea
Let us consider the second term in \C{3one}. Manipulating $\overline\p_{CD}$, we get that
\bea && Y_{IP}Y_{JQ}Y_{CM}Y_{DN}\Big(\overline\p_{CD}\mathcal{B}_1\Big)\Big(\p_{IJ}\mathcal{B}_1\Big)\Big(\p_{MN}\overline\p_{PQ}\mathcal{B}_1\Big) = \frac{25}{32} \mathcal{B}_1 \Big(\mathcal{B}_1^2 - 2 \mathcal{B}_2\Big)\non \\ && -\frac{5}{8}({\rm det}Y)^3 \overline\p_{IJ} \Big[ ({\rm det}Y)^{-3}Y_{IK} Y_{JL}\mathcal{B}_1^2\Big(\p_{KL}\mathcal{B}_1\Big)\Big]\non \\ && + ({\rm det}Y)^3\overline\p_{CD}\Big[ ({\rm det}Y)^{-3}Y_{IP}Y_{JQ}Y_{CM}Y_{DN} \mathcal{B}_1 \Big(\p_{IJ}\mathcal{B}_1\Big)\Big(\p_{MN} \overline\p_{PQ}\mathcal{B}_1\Big)\Big],\eea
where we have used the identity
\bea Y_{IK}Y_{JL}\mathcal{B}_1 \Big(\p_{IJ} \mathcal{B}_1\Big)\Big(\overline\p_{KL} \mathcal{B}_1\Big) = \frac{1}2{}({\rm det}Y)^3 \overline\p_{IJ} \Big[({\rm det}Y)^{-3}Y_{IK}Y_{JL}\mathcal{B}_1^2 \Big(\p_{KL}\mathcal{B}_1\Big)\Big] - \frac{5}{8} \mathcal{B}_1^3,\eea 
and \C{chirel}.

We next consider the third term in \C{3one}. Manipulating $\overline\p_{CD}$, we obtain
\bea \label{long} &&Y_{AK}Y_{BL}Y_{IP}Y_{JQ}\Big(\overline\p_{CD} \mathcal{B}_1\Big)\Big(\p_{IJ}\overline\p_{KL}\mathcal{B}_1\Big)\p_{AB}\Big(Y_{CM}Y_{DN}\p_{MN}\overline\p_{PQ}\mathcal{B}_1\Big) \non \\ &&= ({\rm det}Y)^3 \overline\p_{CD}\Big[ ({\rm det}Y)^{-3}Y_{AK}Y_{BL}Y_{IP}Y_{JQ}\mathcal{B}_1 \Big(\p_{IJ}\overline\p_{KL}\mathcal{B}_1\Big)\p_{AB}\Big(Y_{CM}Y_{DN}\p_{MN}\overline\p_{PQ}\mathcal{B}_1\Big)\Big]\non \\&& -i Y_{BL} Y_{IP} Y_{JQ} \mathcal{B}_1 \Big(\p_{IJ} \overline\p_{KL} \mathcal{B}_1 \Big)\p_{AB} \Big(Y_{AM} Y_{KN} \p_{MN} \overline\p_{PQ} \mathcal{B}_1\Big)\non \\ &&-i Y_{AK} Y_{BL} Y_{JQ} \mathcal{B}_1 \Big(\p_{IJ} \overline\p_{KL} \mathcal{B}_1 \Big)\p_{AB} \Big(Y_{IM} Y_{PN} \p_{MN} \overline\p_{PQ} \mathcal{B}_1\Big)\non \\ &&+i Y_{AK} Y_{BL} Y_{IP} Y_{JQ} \mathcal{B}_1 \Big(\p_{IJ} \overline\p_{KL} \mathcal{B}_1 \Big)\p_{AB} \Big(Y_{MN} \p_{MP} \overline\p_{NQ} \mathcal{B}_1\Big)\non \\&& +i Y_{AK} Y_{BL} Y_{IP} \mathcal{B}_1 \Big(\p_{IK} \overline\p_{CD} \mathcal{B}_1 \Big)\p_{AB} \Big(Y_{CM} Y_{DN} \p_{MN} \overline\p_{PL} \mathcal{B}_1\Big)\non \\ &&-\frac{25}{32} \mathcal{B}_1\mathcal{B}_2 -Y_{AK}Y_{BL}\mathcal{B}_1\p_{AB}\Big(Y_{CM}Y_{DN} \p_{MN} \overline\p_{PQ}\mathcal{B}_1\Big)\overline\p_{KL}\Big(Y_{IP}Y_{JQ}\p_{IJ}\overline\p_{CD}\mathcal{B}_1\Big).\eea
In obtaining \C{long}, at an intermediate step we have manipulated an expression by using
\bea \overline\p_{CD}\p_{AB} \Big(Y_{CM}Y_{DN} \p_{MN} \overline\p_{PQ} \mathcal{B}_1\Big) &= & \p_{AB} \overline\p_{PQ} \Big(Y_{CM} Y_{DN} \p_{MN} \overline\p_{CD} \mathcal{B}_1\Big)+\ldots \non \\ &=& \frac{5}{4} \p_{AB} \overline\p_{PQ} \mathcal{B}_1 +\ldots \eea
to obtain a simplified expression. 

We now consider the last term in \C{long} which we express differently which will be very useful for our purposes. We start with
the expression
\bea \label{long2}\frac{25}{64} \mathcal{B}_1 \Delta \mathcal{B}_2 = Y_{AK}Y_{BL}\mathcal{B}_1\p_{AB} \overline\p_{KL}\Big[Y_{CM}Y_{DN} Y_{IP}Y_{JQ} \Big(\p_{MN} \overline\p_{PQ}\mathcal{B}_1\Big)\Big(\p_{IJ}\overline\p_{CD}\mathcal{B}_1\Big)\Big]\eea
which directly follows from \C{defchi} and \C{chirel}.
We first consider the term on the left hand side of \C{long2} which we manipulate to have the Laplacian acting on $\mathcal{B}_1$. Thus using \C{eigenKZ}, we get that
\bea \label{Long}\frac{25}{64} \mathcal{B}_1 \Delta \mathcal{B}_2 &=& \frac{25}{16} ({\rm det}Y)^3 \p_{AB}\Big[({\rm det}Y)^{-3}Y_{AK}Y_{BL}\mathcal{B}_1\Big(\overline\p_{KL}\mathcal{B}_2\Big)\Big] \non \\ &&- \frac{25}{16} ({\rm det}Y)^3\overline\p_{KL}\Big[ ({\rm det} Y)^{-3}Y_{AK}Y_{BL}\Big(\p_{AB}\mathcal{B}_1\Big)B_2\Big]+\frac{125}{64}\mathcal{B}_1 \mathcal{B}_2.\eea
Next we consider the right hand side of \C{long2}, which gives us
\bea \label{long3}&&\frac{25}{64} \mathcal{B}_1 \Delta \mathcal{B}_2 =  2Y_{AK}Y_{BL}\mathcal{B}_1\p_{AB}\Big(Y_{CM}Y_{DN} \p_{MN} \overline\p_{PQ}\mathcal{B}_1\Big)\overline\p_{KL}\Big(Y_{IP}Y_{JQ}\p_{IJ}\overline\p_{CD}\mathcal{B}_1\Big)\non \\ &&+ 2i  Y_{AI} Y_{BP} Y_{CM} Y_{DN}\mathcal{B}_1 \Big(\p_{MN} \overline\p_{PQ} \mathcal{B}_1\Big)\p_{AB} \Big(Y_{JQ} \p_{IJ} \overline\p_{CD} \mathcal{B}_1\Big)\non \\ &&+ 2i Y_{AP} Y_{CM} Y_{DN} Y_{KQ} \mathcal{B}_1\Big(\p_{MN} \overline\p_{PQ} \mathcal{B}_1\Big)\overline\p_{CD}\Big(Y_{BL} \p_{AB} \overline\p_{KL} \mathcal{B}_1\Big)\non \\ &&-2i Y_{CM} Y_{DN} Y_{IK} Y_{JQ} Y_{LP}  \mathcal{B}_1 \Big(\p_{MN} \overline\p_{PQ} \mathcal{B}_1\Big)\Big(\p_{IJ} \overline\p_{CD} \overline\p_{KL} \mathcal{B}_1\Big)\non \\ &&-2i Y_{AM} Y_{BL} Y_{IP} Y_{JQ} Y_{KN} \Big(\p_{MN} \overline\p_{PQ} \mathcal{B}_1\Big)\Big(\p_{AB} \p_{IJ} \overline\p_{KL} \mathcal{B}_1\Big)+\frac{125}{32}\mathcal{B}_1 \mathcal{B}_2.\eea
In obtaining \C{long3}, at an intermediate step we have used
\bea Y_{AK}Y_{BL}\p_{AB} \overline\p_{KL} \Big(Y_{IP}Y_{JQ} \p_{IJ} \overline\p_{CD} \mathcal{B}_1\Big) &=& Y_{IP}Y_{JQ} \p_{IJ}\overline\p_{CD} \Big(Y_{AK}Y_{BL} \p_{AB}\overline\p_{KL} \mathcal{B}_1\Big)+\ldots \non \\ &=& \frac{5}{4}Y_{IP}Y_{JQ} \p_{IJ}\overline\p_{CD}  \mathcal{B}_1 +\ldots \eea
which simplifies the resulting expression. 

The first term on the right hand side of \C{long3} is precisely of the form of the last term on the right hand side of \C{long}  which we want to express differently. Thus equating \C{Long} and \C{long3}, we get an expression for it in terms of the other terms in these equations, which we substitute in \C{long} along with its complex conjugate. Thus we consider
\bea \label{add}Y_{AK}Y_{BL}Y_{IP}Y_{JQ}\Big(\overline\p_{CD} \mathcal{B}_1\Big)\Big(\p_{IJ}\overline\p_{KL}\mathcal{B}_1\Big)\p_{AB}\Big(Y_{CM}Y_{DN}\p_{MN}\overline\p_{PQ}\mathcal{B}_1\Big)  + c.c. .\eea
Among other contributions, \C{add} contains terms of the form $i\mathcal{B}_1 (\p \overline\p \mathcal{B}_1)\p\Big(\ldots \p \overline\p \mathcal{B}_1 \Big) + c.c.$ schematically, where the ignored factors can involve factors of $Y_{IJ}$. Thus they yield terms of the form $i\mathcal{B}_1 (\p \overline\p \mathcal{B}_1)\p\Big(\p \overline\p \mathcal{B}_1 \Big) + c.c.$ as well as terms where the $\p$ (or $\overline\p$) acts on the factors of $Y_{IJ}$ in $\ldots \p \overline\p \mathcal{B}_1$. The total contribution of all terms of the form  $i\mathcal{B}_1 (\p \overline\p \mathcal{B}_1)\p\Big(\p \overline\p \mathcal{B}_1 \Big) + c.c.$ vanishes, leading to a striking simplification. 

This gives us that
\bea &&Y_{AK}Y_{BL}Y_{IP}Y_{JQ}\Big(\overline\p_{CD} \mathcal{B}_1\Big)\Big(\p_{IJ}\overline\p_{KL}\mathcal{B}_1\Big)\p_{AB}\Big(Y_{CM}Y_{DN}\p_{MN}\overline\p_{PQ}\mathcal{B}_1\Big)  + c.c.  \non \\  &&=({\rm det}Y)^3 \overline\p_{CD}\Big[ ({\rm det}Y)^{-3}Y_{AK}Y_{BL}Y_{IP}Y_{JQ}\mathcal{B}_1 \Big(\p_{IJ}\overline\p_{KL}\mathcal{B}_1\Big)\p_{AB}\Big(Y_{CM}Y_{DN}\p_{MN}\overline\p_{PQ}\mathcal{B}_1\Big)\Big] \non \\ &&-\frac{25}{32} ({\rm det}Y)^3 \p_{AB}\Big[({\rm det}Y)^{-3}Y_{AK}Y_{BL}\mathcal{B}_1\Big(\overline\p_{KL}\mathcal{B}_2\Big)\Big] \non \\ &&+ \frac{25}{32} ({\rm det}Y)^3\overline\p_{KL}\Big[ ({\rm det} Y)^{-3}Y_{AK}Y_{BL}\Big(\p_{AB}\mathcal{B}_1\Big)B_2\Big]-\frac{125}{128}\mathcal{B}_1 \mathcal{B}_2 + c.c..\eea
In the intermediate stages of the analysis, the expression 
\be Y_{IK}Y_{JQ}Y_{ML}Y_{NP}\mathcal{B}_1\Big(\p_{IJ}\overline\p_{KL}\mathcal{B}_1\Big)\Big(\p_{MN}\overline\p_{PQ}\mathcal{B}_1\Big)\ee
arises, which apart from the factor of $\mathcal{B}_1$ involves \C{drop}. However it vanishes in the final answer.  

Thus adding the various contributions, we obtain the expression for $\chi^{(3)}$ given by
\bea \label{eq2}&&\frac{\chi^{(3)}}{32} = ({\rm det}Y)^3 \p_{AB} \Big[ ({\rm det}Y)^{-3} Y_{AK} Y_{BL}Y_{IP}Y_{JQ}Y_{CM}Y_{DN} \Big(\overline\p_{CD} \mathcal{B}_1\Big)\Big(\p_{IJ} \overline\p_{KL} \mathcal{B}_1\Big) \non \\ &&\times \Big(\p_{MN}\overline\p_{PQ}\mathcal{B}_1\Big)\Big]-({\rm det}Y)^3 \overline\p_{CD}\Big[ ({\rm det}Y)^{-3}Y_{AK}Y_{BL}Y_{IP}Y_{JQ}\mathcal{B}_1 \Big(\p_{IJ}\overline\p_{KL}\mathcal{B}_1\Big)\non \\ &&\times \p_{AB}\Big(Y_{CM}Y_{DN}\p_{MN}\overline\p_{PQ}\mathcal{B}_1\Big)\Big] -\frac{5}{4} ({\rm det}Y)^3\overline\p_{CD}\Big[ ({\rm det}Y)^{-3}Y_{IP}Y_{JQ}Y_{CM}Y_{DN} \mathcal{B}_1 \Big(\p_{IJ}\mathcal{B}_1\Big)\non \\ &&\times \Big(\p_{MN} \overline\p_{PQ}\mathcal{B}_1\Big)\Big] +\frac{25}{32} ({\rm det}Y)^3 \p_{AB}\Big[({\rm det}Y)^{-3}Y_{AK}Y_{BL}\mathcal{B}_1\Big(\overline\p_{KL}\mathcal{B}_2\Big)\Big] \non \\ &&+\frac{25}{32}({\rm det}Y)^3 \overline\p_{IJ} \Big[ ({\rm det}Y)^{-3}Y_{IK} Y_{JL}\mathcal{B}_1^2\Big(\p_{KL}\mathcal{B}_1\Big)\Big]\non \\  &&- \frac{25}{32} ({\rm det}Y)^3\overline\p_{KL}\Big[ ({\rm det} Y)^{-3}Y_{AK}Y_{BL}\Big(\p_{AB}\mathcal{B}_1\Big)\mathcal{B}_2\Big]-\frac{125}{128} \mathcal{B}_1 \Big(\mathcal{B}_1^2 - 3 \mathcal{B}_2\Big)+c.c..\eea

We now equate the two expressions \C{eq} and \C{eq2} which have been obtained by calculating $\chi^{(3)}$ in two different ways. The term involving $\mathcal{B}_1(\mathcal{B}_1^2-3\mathcal{B}_2)$ cancels leading to a reduction in the number of string invariants that appear in the final expression. We get that

\bea \label{simplE}&&-\frac{125\mathcal{B}_9}{32({\rm det}Y)^3} = \p_{AB} \Big[ ({\rm det}Y)^{-3} Y_{AK} Y_{BL}Y_{IP}Y_{JQ}Y_{CM}Y_{DN} \Big(\overline\p_{CD} \mathcal{B}_1\Big)\Big(\p_{IJ} \overline\p_{KL} \mathcal{B}_1\Big) \non \\ &&\times \Big(\p_{MN}\overline\p_{PQ}\mathcal{B}_1\Big)\Big]- \overline\p_{CD}\Big[ ({\rm det}Y)^{-3}Y_{AK}Y_{BL}Y_{IP}Y_{JQ}\mathcal{B}_1 \Big(\p_{IJ}\overline\p_{KL}\mathcal{B}_1\Big)\non \\ && \times \p_{AB}\Big(Y_{CM}Y_{DN}\p_{MN}\overline\p_{PQ}\mathcal{B}_1\Big)\Big]  -\frac{5}{4} \overline\p_{CD}\Big[ ({\rm det}Y)^{-3}Y_{IP}Y_{JQ}Y_{CM}Y_{DN} \mathcal{B}_1 \Big(\p_{IJ}\mathcal{B}_1\Big)\non \\ && \times \Big(\p_{MN} \overline\p_{PQ}\mathcal{B}_1\Big)\Big] +\frac{25}{32}  \p_{AB}\Big[({\rm det}Y)^{-3}Y_{AK}Y_{BL}\mathcal{B}_1\Big(\overline\p_{KL}\mathcal{B}_2\Big)\Big] \non \\ &&+\frac{25}{32} \overline\p_{IJ} \Big[ ({\rm det}Y)^{-3}Y_{IK} Y_{JL}\mathcal{B}_1^2\Big(\p_{KL}\mathcal{B}_1\Big)\Big]- \frac{25}{32} \overline\p_{KL}\Big[ ({\rm det} Y)^{-3}Y_{AK}Y_{BL}\Big(\p_{AB}\mathcal{B}_1\Big)\mathcal{B}_2\Big]+c.c..\non \\ \eea
Thus we see that the string invariant $\mathcal{B}_9$ is completely determined by the KZ invariant $\mathcal{B}_1$ and $\mathcal{B}_2$, which in turn is determined by the KZ invariant using \C{simplify1}. Also $\mathcal{B}_9/({\rm det}Y)^3$ is a total derivative on moduli space which will be very useful for our purposes. 

To simplify \C{simplE}, along with \C{YY} we use
\bea \overline\p_{CD} \Big[ \frac{\mathcal{B}_1^2}{({\rm det}Y)^3}Y_{AP}Y_{BQ} \p_{AB}\Big(Y_{CM}Y_{DN}\p_{MN}\overline\p_{PQ} \mathcal{B}_1\Big)\Big] = \frac{5}{4} \overline\p_{CD} \Big[ \frac{\mathcal{B}_1^2}{({\rm det}Y)^3}Y_{CM}Y_{DN}\p_{MN}\mathcal{B}_1\Big]. \non \\ \eea
This gives us that
\bea \label{simpLe}&&-\frac{125\mathcal{B}_9}{32({\rm det}Y)^3} = \frac{25}{32} \overline\p_{KL}\Big[ \frac{Y_{IK}Y_{JL}}{({\rm det}Y)^3}\Big(\mathcal{B}_1 \p_{IJ}\mathcal{B}_2 - \mathcal{B}_2\p_{IJ}\mathcal{B}_1 - \frac{1}{3} \p_{IJ} \mathcal{B}_1^3\Big)\Big]\non \\ &&-\epsilon_{AI}\epsilon_{BJ}\epsilon_{KP}\epsilon_{LQ} \overline\p_{CD}\Big[\frac{\mathcal{B}_1}{{\rm det}Y}\Big(\p_{IJ}\overline\p_{KL}\mathcal{B}_1\Big)\p_{AB}\Big(Y_{CM}Y_{DN}\p_{MN}\overline\p_{PQ}\mathcal{B}_1\Big)\Big]\non \\ &&-2\epsilon_{AI}\epsilon_{KP}\overline\p_{CD}\Big[\frac{\mathcal{B}_1}{({\rm det}Y)^2} Y_{JL}Y_{BQ}\Big(\p_{IJ}\overline\p_{KL}\mathcal{B}_1\Big)\p_{AB}\Big(Y_{CM}Y_{DN}\p_{MN}\overline\p_{PQ}\mathcal{B}_1\Big)\Big]\non \\ &&+\epsilon_{AI}\epsilon_{BJ}\epsilon_{KP}\epsilon_{LQ} \p_{AB}\Big[\frac{\overline\p_{CD}\mathcal{B}_1}{{\rm det}Y}\Big(\p_{IJ}\overline\p_{KL}\mathcal{B}_1\Big)\Big(Y_{CM}Y_{DN}\p_{MN}\overline\p_{PQ}\mathcal{B}_1\Big)\Big]\non \\  &&+2\epsilon_{AI}\epsilon_{KP}\p_{AB}\Big[\frac{\overline\p_{CD}\mathcal{B}_1}{({\rm det}Y)^2} Y_{JL}Y_{BQ}\Big(\p_{IJ}\overline\p_{KL}\mathcal{B}_1\Big)\Big(Y_{CM}Y_{DN}\p_{MN}\overline\p_{PQ}\mathcal{B}_1\Big)\Big]+c.c..\eea

\section{Integrating simple string invariants over moduli space}

We now consider integrating some simple string invariants we have discussed above over moduli space. Such integrals are of the form
\be \label{Sp}\int_{\mathcal{F}_2} d\mu \frac{\mathcal{B}}{({\rm det}Y)^3}.\ee
In \C{Sp}, $\mathcal{B}$ is constructed out of string invariants, and the integral is over $\mathcal{F}_2$, the fundamental domain of $Sp(4,\mathbb{Z})$.
The $Sp(4,\mathbb{Z})$ invariant measure $d\mu/({\rm det}Y)^3$ involves 
\be d\mu = \prod_{I\leq J} 2 d({\rm Re} \Omega_{IJ}) \wedge d ({\rm Im}\Omega_{IJ}).\ee

Such integrals in \C{Sp} for generic $\mathcal{B}$ are difficult to evaluate, as they involve data all over moduli space, and also given the involved structure of $\mathcal{F}_2$. However, if the integral can be reduced to a boundary term in moduli space,  its evaluation becomes considerably simpler as it only involves boundary data. For example, this method of evaluating the integral facilitated the evaluation of the integral of the KZ invariant over moduli space~\cite{DHoker:2014oxd}.

We now consider \C{Sp} when $\mathcal{B}$ is $\mathcal{B}_2 - \mathcal{B}_1^2$, and also when it is $\mathcal{B}_9$.  Now from \C{g1} and \C{simpLe}, we see that both these integrals reduce to boundary terms in moduli space. To evaluate them, we first briefly describe the structure of the boundary of moduli space, and the asymptotic expansions of the relevant string invariants, and then proceed to evaluate the integrals.

To analyze the boundary structure, we parametrize the period matrix $\Omega$ as
\be \label{parap}\Omega= \left( \begin{array}{cc} \tau & v \\ v & \sigma \end{array} \right).\ee
The boundary of moduli space involves contributions from the separating and the non--separating nodes, as well as their intersection. 

The separating node is obtained from \C{parap} by taking $v \rightarrow 0$, while keeping $\tau, \sigma$ fixed. At this node, an $SL(2,\mathbb{Z})_\tau \times SL(2,\mathbb{Z})_\sigma$ subgroup of $Sp(4,\mathbb{Z})$ survives with the action
\be \label{sepexp}v \rightarrow \frac{v}{(c\tau+d)(c'\sigma +d')}, \quad \tau \rightarrow \frac{a\tau+b}{c\tau+d}, \quad \sigma \rightarrow \frac{a'\sigma +b'}{c'\sigma +d'},\ee 
where $a,b,c,d,a',b',c',d' \in \mathbb{Z}$ and $ad-bc = a'd' -b'c' =1$. Thus $\tau$ and $\sigma$ parametrize the complex structure moduli of the resulting tori.  

The non--separating node is obtained by taking $\sigma \rightarrow i\infty$, while keeping $\tau, v$ fixed\footnote{Another contribution comes from taking $\tau \rightarrow i\infty$, while keeping $\sigma, v$ fixed. These two contributions are simply related by $\tau \leftrightarrow \sigma$ exchange, and hence we focus on only one of them.}. At this node, an $SL(2,\mathbb{Z})_\tau$ subgroup of $Sp(4,\mathbb{Z})$ survives whose action on $v, \tau$ and $\sigma$ is given by~\cite{Moore:1986rh,DHoker:2017pvk,DHoker:2018mys}
\be v \rightarrow \frac{v}{(c\tau+d)}, \quad \tau \rightarrow \frac{a\tau+b}{c\tau+d},\quad \sigma \rightarrow \sigma - \frac{cv^2}{c\tau+d},\ee 
where $a,b,c,d \in \mathbb{Z}$ and $ad-bc=1$. At this node, $v$ parametrizes the coordinate on the torus with complex structure $\tau$, and thus
\be -\frac{1}{2} \leq v_1 \leq \frac{1}{2}, \quad 0 \leq v_2 \leq \tau_2.\ee
Also $\sigma_2$ along with $v_2$ and $\tau_2$ forms the $SL(2,\mathbb{Z})_\tau$ invariant quantity
\be \label{deft}t= \sigma_2 - \frac{v_2^2}{\tau_2}\ee
which we shall use later.

We now consider the asymptotic expansion of various quantities around these nodes that will be relevant for our analysis. To start with, ${\rm det} Y$ behaves as
\be \label{det1} {\rm det} Y = \tau_2  \sigma_2 +O(v_2^2)\ee
at the separating node, and as
\be \label{det2}{\rm det} Y = \tau_2\sigma_2 + O(\sigma_2^0)\ee
at the non--separating node. We now state the expressions for the asymptotic expansions of the string invariants $\mathcal{B}_1$ and $\mathcal{B}_2$~\cite{Wentworth,Jong,DHoker:2013fcx,Pioline:2015qha,DHoker:2017pvk,DHoker:2018mys}.

The asymptotic expansions of $\mathcal{B}_1$ and $\mathcal{B}_2$ around the separating node are given by
\bea \label{sep}\mathcal{B}_1 &=& 4{\rm ln}\vert \lambda \vert + O(\vert \lambda \vert), \non \\ \mathcal{B}_2 &=& 16 {\rm ln}^2 \vert \lambda \vert +O(\vert \lambda \vert),\eea
where\footnote{Note that $\vert \lambda \vert$ is $SL(2,\mathbb{Z})_\tau \times SL(2,\mathbb{Z})_\sigma$ invariant under the transformations \C{sepexp}.}
\be \lambda = 2\pi v \eta^2 (\tau) \eta^2 (\sigma),\ee
and $\eta (\tau)$ is the Dedekind eta function.

On the other hand, the asymptotic expansions around the non--separating node are given by\footnote{In the asymptotic expansion of $\mathcal{B}_2$, apart from terms that are $O(e^{-t})$, we have also ignored the $O(t^0)$, $O(t^{-1})$ and $O(t^{-2})$ terms~\cite{DHoker:2017pvk,DHoker:2018mys}, as they are not relevant for our analysis. }
\bea \label{nonsep}\mathcal{B}_1 &=&  -\frac{2\pi t}{3} -2 g(v) -\frac{5F_2 (v)}{\pi t} +O(e^{-t}), \non \\ 
\mathcal{B}_2 &=& \frac{4\pi^2t^2}{9}+\frac{8\pi t g(v)}{3} +O(t^0).\eea

In \C{nonsep}, $g(v)$ is the genus one Green function given by
\be \label{Green}g(v)\equiv g(v;\tau) = \sum_{(m,n) \neq (0,0)}\frac{\tau_2}{\pi\vert m+n\tau\vert^2} e^{2\pi i(my-nx)},\ee
where we have parametrized $v$ as
\be \label{param}v = x+\tau y,\ee
with $x,y \in (0,1]$. 
The Green function is single--valued and doubly periodic on the torus. Thus
we have that
\be \label{zerogreen}\int_{\S} d^2 z g (z)= 0, \quad \int_{\S} d^2 z \p_z \overline\p_z g (z)=0,\ee
which follows from \C{Green}, where $\Sigma$ is the toroidal worldsheet.

In \C{nonsep}, we also have that\footnote{Note that $g_2 (v)$ and $E_2$ are $SL(2,\mathbb{Z})_\tau$ invariant.}
\be F_2 (v) = E_2 - g_2 (v) ,\ee
where $g_2 (v)$ is the iterated Green function defined by
\be g_2 (v) \equiv g_2 (v;\tau) =  \int_{\Sigma} \frac{d^2 z}{2\tau_2} g(v-z;\tau)g(z;\tau)\ee
and
\be E_2 \equiv E_2 (\tau) = g_2 (0)\ee
is the non--holomorphic Eisenstein series. 

Now $g(v)$, $g_2 (v)$ and $E_2$ satisfy the differential equations
\bea \label{manyeqn}&& \Delta_\tau g(v) =0, \quad \Delta_v g(v) = -8\pi\tau_2 \delta^2 (v) + 4\pi, \non \\ && \Delta_\tau g_2 (v) = 2 g_2 (v), \quad \Delta_v g_2 (v) = -4\pi g(v), \quad \Delta_\tau E_2 = 2 E_2,\eea
which are useful in our analysis. In \C{manyeqn}  we have defined the $SL(2,\mathbb{Z})_\tau$ invariant operators\footnote{The $\tau$ derivative is taken at constant $x, y$ in \C{param}, and not at constant $v$.}
\be \Delta_\tau = 4\tau_2^2 \p_\tau \overline\p_\tau , \quad \Delta_v = 4\tau_2 \p_v \overline\p_v ,\ee
while the delta function is normalized to satisfy $\int_{\S} d^2 z \delta^2 (z)=1$. 

In order to evaluate the integrals of $\mathcal{B}_2 - \mathcal{B}_1^2$ and $\mathcal{B}_9$ over moduli space, we shall think of the boundary contributions as limits of contributions in the bulk~\cite{DHoker:2014oxd}. To see the structure, consider the integral (or its complex conjugate) given by\footnote{We define
\be \p_\tau = \frac{\p}{\p\tau}, \quad \p_\sigma = \frac{\p}{\p\sigma}, \quad \p_v= \frac{\p}{\p v}\ee
and similarly for its complex conjugates.} 
\bea \label{boundary}\int_{\mathcal{F}_2} d\mu \overline\p_{IJ} \Psi_{IJ} = \int_{\mathcal{F}_2} d\mu\Big[\overline\p_{\tau} \Psi_{11} +\overline\p_{\s} \Psi_{22} +\frac{1}{2}\overline\p_v \Big(\Psi_{12} +\Psi_{21}\Big)\Big]\eea
where we have used \C{pdef}.
The first two terms receive contributions from the non--separating node while the remaining terms receive contributions from the separating node, and hence the integral is entirely determined by the asymptotic expansions of $\Psi_{IJ}$ around these nodes. In this limiting procedure, the contribution from the separating node is evaluated in the complex $v$ plane as $\vert v \vert = R \rightarrow 0$. Hence this is an integral in the complex $v$ plane on a circle around the origin with vanishing radius.  On the other hand, the contribution from the non--separating node $\s_2 \rightarrow \infty$ is evaluated as $t= L\rightarrow \infty$ using \C{deft} in an $SL(2,\mathbb{Z})_\tau$ invariant way\footnote{This essentially reduces to neglecting $v_2 = y \tau_2$ contributions in the final answer that result from various expressions involving $Y_{IJ}$. After taking this limit, what remains is an $SL(2,\mathbb{Z})_\tau$ invariant integral over $v$ and $\tau$.}.

We shall see that there are no divergences in this limit in the two cases we consider. For $\mathcal{B}_2 - \mathcal{B}_1^2$, it follows from \C{sep} since $\mathcal{B}_2 - \mathcal{B}_1^2 \sim O(\vert \lambda \vert)$ at the separating node, and from \C{nonsep} as $\mathcal{B}_2 - \mathcal{B}_1^2 \sim O(t)$ at the non--separating node. Thus they lead to absolutely convergent integrals in \C{Sp} on simply using \C{det1} and \C{det2}. For $\mathcal{B}_9$, a similar conclusion should follow from its asymptotic expansions around the various nodes. 

Before we proceed to calculate these integrals, let us very briefly consider the case when $\mathcal{B} = \mathcal{B}_1$~\cite{DHoker:2014oxd} in \C{Sp}. Using \C{eigenKZ}, we have that
\be \int_{\mathcal{F}_2} d\mu \frac{\mathcal{B}_1}{({\rm det}Y)^3} =\frac{4}{5} \int_{\mathcal{F}_2} d\mu \p_{IJ} \Big[ \frac{Y_{IK}Y_{JL}}{({\rm det}Y)^3}\overline\p_{KL}\mathcal{B}_1\Big],\ee
hence reducing to an integral over the boundary of moduli space.
Now from \C{nonsep}, we see that the contribution from the non--separating node vanishes. On the other hand, from \C{sep} we see that the contribution from the separating node is of the form
\be \label{nonvan}\int_{\mathcal{F}_1} \frac{d^2\tau}{\tau_2^2} \int_{\mathcal{F}_1}\frac{d^2\s}{\s_2^2}\oint \frac{d\overline{v}}{\overline{v}},\ee
where $\mathcal{F}_1$ is the fundamental domain of $SL(2,\mathbb{Z})$ and the contour has been mentioned above. Thus \C{nonvan} is non--vanishing because of the presence of the simple pole in the contour integral. We shall see this crucial feature is absent in the integrals which we now analyze.   

\subsection{Integral involving simple string invariants with two links}

We first consider the integral 
\be \int_{\mathcal{F}_2} d\mu \frac{\Big(\mathcal{B}_2 - \mathcal{B}_1^2\Big)}{({\rm det}Y)^3}\ee
over moduli space, which using \C{g1} reduces to a boundary term which we evaluate based on the discussion above. Its evaluation requires the asymptotic expansions of $\mathcal{B}_1$ given by \C{sep} and \C{nonsep}. 

First let us consider the contribution to the integral from the separating node, where the only potentially non--vanishing contributions arise from the terms given in \C{sep} in the limit $R \rightarrow 0$. The first term in \C{g1} gives us 
\bea \frac{1}{4} \overline\p_v \Big[ \frac{\Big(\p_v \mathcal{B}_1\Big) \Big(\p_v \overline\p_v \mathcal{B}_1\Big)}{\tau_2 \sigma_2}\Big],\eea
which using $\p_v \overline\p_v \mathcal{B}_1 \sim \delta^2 (v)$, yields a divergent contribution of the form $\delta^2(v)/v$ on the boundary using \C{boundary}. However the second term in \C{g1} produces a cancelling contribution, and so the total contribution at the separating node vanishes. 

We next consider the contribution to the integral from the non--separating node using \C{nonsep} as the ignored terms do not contribute. While the first term in \C{g1} does not contribute, the second term gives us
\be \label{pre1}-\frac{\pi}{6} \p_t \Big[\frac{\p_v \overline\p_v g(v)}{\tau_2^2}\Big].\ee  
Thus using \C{boundary}, it yields an $SL(2,\mathbb{Z})_\tau$ invariant contribution proportional to\footnote{The integral over $\sigma_1$ simply yields
\be \int_0^1 d\s_1 = 1.\ee
The $\sigma_1$ dependence in the asymptotic expansions of the string invariants comes from terms of the form $e^{2\pi i \sigma}$ which are exponentially suppressed for large $t$ and do not contribute to the answer.}
\be \label{zerO}\int_{\mathcal{F}_1} \frac{d^2\tau}{\tau_2^2}\int_{\S} \frac{d^2v}{\tau_2} \Delta_v g(v)\ee
in the final expression.  Using \C{zerogreen}, the integral over $\S$ vanishes. Hence there is no contribution from the non--separating node. Thus this yields
\be \int_{\mathcal{F}_2} d\mu \frac{\Big(\mathcal{B}_2 - \mathcal{B}_1^2\Big)}{({\rm det}Y)^3} =0\ee
leading to a vanishing integral.

\subsection{Integral involving a simple string invariant with three links}

We next consider the integral 
\be \int_{\mathcal{F}_2} d\mu \frac{\mathcal{B}_9 }{({\rm det}Y)^3}\ee
over moduli space, which using \C{simpLe} again reduces to a boundary term which we now evaluate. Note that its evaluation requires only the asymptotic expansions of $\mathcal{B}_1$ and $\mathcal{B}_2$ given by \C{sep} and \C{nonsep}, and does not require any information about $\mathcal{B}_9$, hence simplifying the analysis considerably.

To begin with, consider the contribution to the integral from the separating node, where the potentially non--vanishing contributions arise from \C{sep}. 
Now using
\be \label{reduce}\mathcal{B}_1 \p_{IJ}\mathcal{B}_2 - \mathcal{B}_2\p_{IJ}\mathcal{B}_1 - \frac{1}{3} \p_{IJ} \mathcal{B}_1^3 = \Big(\mathcal{B}_1^2 - \mathcal{B}_2\Big)\p_{IJ} \mathcal{B}_1 +\mathcal{B}_1 \p_{IJ} \Big(\mathcal{B}_2 - \mathcal{B}_1^2\Big)\ee
and that $\mathcal{B}_2 - \mathcal{B}_1^2 \sim O(\vert \lambda \vert)$ at this node, we see that the contribution from the first term in \C{simpLe} vanishes. 

The second term contributes
\bea -\frac{1}{8} \overline\p_v\Big[ \mathcal{B}_1\Big(\p_v \overline\p_v \mathcal{B}_1\Big)\p_v \Big(\p_v \overline\p_v \mathcal{B}_1\Big)\Big]\eea
which yields a contribution involving $\delta^2(v) \p_v \delta^2 (v){\rm ln}\vert \lambda \vert$ on the boundary of moduli space. However, the third term produces a cancelling contribution.  

The fourth term contributes
\bea \frac{1}{8} \p_v\Big[ \frac{\overline\p_v\mathcal{B}_1}{\tau_2 \sigma_2}\Big(\p_v \overline\p_v \mathcal{B}_1\Big)^2 \Big]\eea
which yields a contribution involving $(\delta^2(v))^2/\overline{v}$ on the boundary of moduli space. The fifth term produces a cancelling contribution. Thus the total contribution at the separating node vanishes. 

We next consider the contribution to the integral from the non--separating node. As $t= L \rightarrow \infty$, the relevant contributions from \C{simpLe} to the final expression must be $SL(2,\mathbb{Z})_\tau$ invariant, and hence we focus on them. 

The contribution from the first term in \C{simpLe} is of the form
\be  \p_t \Big[ \frac{1}{t\tau_2^3} \Big(\mathcal{B}_1 \p_t \mathcal{B}_2 - \mathcal{B}_2 \p_t \mathcal{B}_1 - \frac{1}{3}\p_t \mathcal{B}_1^3 \Big)\Big],\ee
resulting in 
\be \frac{1}{t}\int_{\mathcal{F}_1} \frac{d^2\tau}{\tau_2^2}\int_{\S} \frac{d^2v}{\tau_2} \Big[\Big(\mathcal{B}_1^2 - \mathcal{B}_2\Big)\p_t \mathcal{B}_1 +\mathcal{B}_1 \p_t \Big(\mathcal{B}_2 - \mathcal{B}_1^2\Big)\Big]\Big\vert_{t=L\rightarrow \infty}\ee
in the final expression. From \C{nonsep} we see that $\mathcal{B}_2 - \mathcal{B}_1^2 \sim O(t^0)$ and hence the first term in \C{simpLe} does not contribute. 

The $SL(2,\mathbb{Z})_\tau$ invariant contributions from the second term in \C{simpLe} arise from
\be \label{2t}-\frac{i}{8}\p_t \Big[ \frac{\mathcal{B}_1}{t\tau_2} \Big(\p_v\overline\p_v \mathcal{B}_1\Big) \p_v\Big(Y_{2M} Y_{2N} \p_{MN} \overline\p_v \mathcal{B}_1\Big)\Big],\ee
which is exactly cancelled by a contribution coming from the third term in \C{simpLe}. This is much like the analysis for the separating node where competing contributions cancel. However, \C{2t} produces only vanishing contributions by itself and one need not consider other terms. This is like the analysis of the non--separating node in the previous section. To see this, from \C{nonsep} we have that 
\bea \label{V1}\frac{\mathcal{B}_1}{t\tau_2} \Big(\p_v\overline\p_v \mathcal{B}_1\Big) \sim \frac{\p_v\overline\p_v g(v)}{\tau_2} +\frac{1}{t\tau_2}\Big(\frac{g(v)}{\tau_2}+ g(v)\p_v\overline\p_v g(v)\Big) + O(t^{-2}),\eea
where we have used
\be \Delta_v F_2 (v) = 4\pi g(v).\ee        
We also have that
\be \label{V2} \p_v\Big(Y_{2M} Y_{2N} \p_{MN} \overline\p_v \mathcal{B}_1\Big)\sim \frac{t}{\tau_2}\ee
as the $O(t^0)$ contribution cancels.  

Thus from \C{V1}, \C{V2} and \C{2t}, we get a contribution of the form 
\be \label{C1}\p_t \Big[\frac{t \p_v \overline\p_v g(v)}{\tau_2^2}\Big] \ee 
yielding a potentially linearly divergent contribution
\be L \int_{\mathcal{F}_1} \frac{d^2\tau}{\tau_2^2}\int_{\S} \frac{d^2v}{\tau_2} \Delta_v g(v)\ee
at the boundary.
However, the integral is the same as \C{zerO} and vanishes. 

We also get finite contributions as $L \rightarrow \infty$. One of them is of the form
\be \label{C2}\p_t \Big[ \frac{g(v)}{\tau_2^3}\Big],\ee
which yields the boundary term
\be \int_{\mathcal{F}_1}\frac{d^2\tau}{\tau_2^2}\int_{\S} \frac{d^2 v}{\tau_2} g(v),\ee 
which vanishes using \C{zerogreen}. The other finite contribution is of the form
\be \label{C3}\p_t \Big[ \frac{g(v) \Delta_v g(v)}{\tau_2^3}\Big]\ee
leading to the boundary term
\be \int_{\mathcal{F}_1}\frac{d^2\tau}{\tau_2^2}\int_{\S} \frac{d^2 v}{\tau_2} g(v)\Delta_v g(v).\ee
However on using \C{manyeqn}, and setting $g(0)=0$\footnote{Coincident Green functions, resulting from colliding vertex operators, are not in the moduli space of these graphs as they produce other local operators using the operator product expansion, the propagation of which leads to kinematic poles, rather than contact interactions in the amplitude. These cannot be seen by a naive perturbative expansion in $\alpha'$ by keeping a finite number of terms. In fact, this follows from an analysis of the structure of the Koba--Nielsen factor in the string amplitude using the cancelled propagator argument (see \cite{DHoker:2020aex}, for example, for a recent discussion). }, we see this contribution vanishes using \C{zerogreen}. 

Proceeding similarly, we see that the remaining terms in \C{simpLe} do not give any additional non--vanishing contributions.
Thus the total contribution from the non--separating node vanishes. 

Hence this leads to the vanishing integral
\be \int_{\mathcal{F}_2} d\mu \frac{\mathcal{B}_9 }{({\rm det}Y)^3} =0\ee
over moduli space.

Thus we see that manipulating the expressions $\chi^{(i)} (\Omega,\overline\Omega)$ by evaluating them in two different ways leads to results for integrals of some simple string invariants over moduli space. We expect that generalizing this analysis for graphs with more links will prove useful in evaluating various integrals over moduli space by reducing them to boundary terms, which depend only on asymptotic data.      

\appendix

\section{Relations involving the graphs $\mathcal{B}_5, \mathcal{B}_6, \mathcal{B}_7, \mathcal{B}_8, \mathcal{B}_9$, $\mathcal{B}_{10}$ and $\mathcal{B}_{11}$}

We now obtain various relations involving the graphs $\mathcal{B}_5, \mathcal{B}_6, \mathcal{B}_7, \mathcal{B}_8, \mathcal{B}_9$, $\mathcal{B}_{10}$ and $\mathcal{B}_{11}$ defined by \C{defB} and \C{defB11} that arise in the analysis of $\chi^{(3)}$. They prove to be very useful in simplifying \C{finval}.

To start with, we define
\be \Delta(z_i,z_j) = \epsilon_{IJ} \omega_I (z_i) \omega_J (z_j),\ee
which satisfies the identity~\cite{DHoker:2020tcq}
\be \label{Rel}\omega_I (z_i)\Delta (z_j,z_k)+ \omega_I (z_k)\Delta (z_i,z_j)+\omega_I (z_j)\Delta (z_k,z_i) =0.\ee

In our analysis, we also use the identity 
\be \label{relDel}({\rm det}Y)^{-1} \Delta (z_i,z_j) \overline{\Delta(z_k,z_l)} = (z_i,\overline{z_k})(z_j,\overline{z_l}) - (z_i,\overline{z_l})(z_j,\overline{z_k}).\ee

First let us consider the graph $\mathcal{B}_5$. Using \C{relDel} with $z_i =z_3$, $z_j = z_4$, $z_k =z_1$ and $z_l=z_2$, we get that\footnote{We often use \C{zero} in our analysis.}
\bea \mathcal{B}_5 - \mathcal{B}_6 &=& \int_{\S^6} \prod_{i=1}^6 d^2 z_i G(z_1,z_2)G(z_3,z_4)G(z_5,z_6)(z_5,\overline{z_6})(z_6,\overline{z_3})(z_2,\overline{z_4})(z_1,\overline{z_5})\non \\ &&\times \frac{\Delta(z_3,z_4)\overline{\Delta(z_1,z_2)}}{{\rm det}Y}.\eea 
For the term on the right hand side, we use \C{Rel} with $z_i = z_2$, $z_j = z_3$ and $z_k = z_4$ as well as \C{relDel} to get the relation
\be \label{rel56}\mathcal{B}_5 = \mathcal{B}_6 +\mathcal{B}_{11}.\ee

Let us again consider the graph $\mathcal{B}_5$. Using \C{relDel} with $z_i =z_6$, $z_j = z_4$, $z_k =z_3$ and $z_l=z_2$ instead, we obtain
\bea \mathcal{B}_5 - \mathcal{B}_7 &=& \int_{\S^6} \prod_{i=1}^6 d^2 z_i G(z_1,z_2)G(z_3,z_4)G(z_5,z_6)(z_5,\overline{z_6})(z_3,\overline{z_1})(z_2,\overline{z_4})(z_1,\overline{z_5})\non \\ &&\times \frac{\Delta(z_6,z_4)\overline{\Delta(z_3,z_2)}}{{\rm det}Y}.\eea 
For the term on the right hand side, using \C{Rel} with $z_i = z_2$, $z_j = z_6$ and $z_k = z_4$ as well as \C{relDel} we get that
\be \mathcal{B}_5 = \mathcal{B}_7 +\mathcal{B}_6,\ee
and hence 
\be \label{rel711}\mathcal{B}_7 = \mathcal{B}_{11}\ee
which follows from \C{rel56}. 

We next consider the graph $\mathcal{B}_7$. Using \C{relDel} with $z_i =z_4$, $z_j = z_3$, $z_k =z_2$ and $z_l=z_5$, yields 
\bea \mathcal{B}_7 - \mathcal{B}_8 &=& \int_{\S^6} \prod_{i=1}^6 d^2 z_i G(z_1,z_2)G(z_3,z_4)G(z_5,z_6)(z_1,\overline{z_3})(z_5,\overline{z_6})(z_6,\overline{z_4})(z_2,\overline{z_1})\non \\ &&\times \frac{\Delta(z_4,z_3)\overline{\Delta(z_2,z_5)}}{{\rm det}Y}.\eea 
For the term on the right hand side, we use \C{Rel} with $z_i = z_6$, $z_j = z_4$ and $z_k = z_3$ as well as \C{relDel} to obtain the relation
\be \label{rel78}\mathcal{B}_7 = \mathcal{B}_8 -\mathcal{B}_{11} +\mathcal{B}_1 \mathcal{B}_4.\ee

Finally let us consider the graph $\mathcal{B}_9$. Using \C{relDel} with $z_i =z_3$, $z_j = z_4$, $z_k =z_1$ and $z_l=z_2$, we get that 
\bea \mathcal{B}_9 - \mathcal{B}_{10} &=& \int_{\S^6} \prod_{i=1}^6 d^2 z_i G(z_1,z_2)G(z_3,z_4)G(z_5,z_6)(z_1,\overline{z_5})(z_5,\overline{z_3})(z_2,\overline{z_6})(z_6,\overline{z_4})\non \\ &&\times \frac{\Delta(z_3,z_4)\overline{\Delta(z_1,z_2)}}{{\rm det}Y}.\eea 
For the term on the right hand side, using \C{Rel} with $z_i = z_6$, $z_j = z_3$ and $z_k = z_4$ as well as \C{relDel} we get that
\be \mathcal{B}_9 = \mathcal{B}_{10} -\mathcal{B}_5 +\mathcal{B}_6 = \mathcal{B}_{10} -\mathcal{B}_{11}\ee
on using \C{rel56}.

\begin{figure}[ht]
\begin{center}
\[
\mbox{\begin{picture}(200,130)(0,0)
\includegraphics[scale=.75]{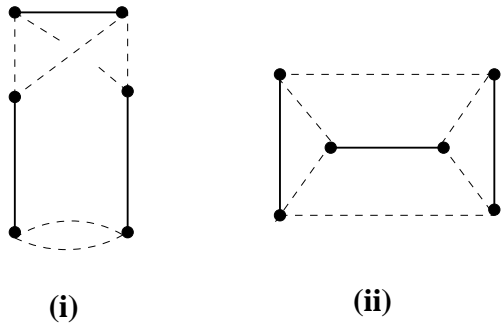}
\end{picture}}
\]
\caption{The string invariants (i) $\mathcal{B}_{12}$ and (ii) $\mathcal{B}_{14}$}
\end{center}
\end{figure}

To obtain more relations between these graphs, we introduce the graphs $\mathcal{B}_{12}$, $\mathcal{B}_{13}$ and $\mathcal{B}_{14}$ defined by
\bea \label{defBmore} \mathcal{B}_{12} &=& \int_{\S^6} \prod_{i=1}^6 d^2 z_i G(z_1,z_2)G(z_3,z_4)G(z_5,z_6) (z_2,\overline{z_3})(z_3,\overline{z_1})(z_1,\overline{z_6})(z_6,\overline{z_2})P(z_4,z_5),\non \\ \mathcal{B}_{13} &=& \int_{\S^6} \prod_{i=1}^6 d^2 z_i G(z_1,z_2)G(z_3,z_4)G(z_5,z_6) (z_1,\overline{z_3})(z_3,\overline{z_5})(z_5,\overline{z_1})(z_6,\overline{z_4})(z_4,\overline{z_2})(z_2,\overline{z_6}), \non \\ \mathcal{B}_{14} &=& \int_{\S^6} \prod_{i=1}^6 d^2 z_i G(z_1,z_2)G(z_3,z_4)G(z_5,z_6) (z_1,\overline{z_3})(z_3,\overline{z_2})(z_2,\overline{z_6})(z_6,\overline{z_4})(z_4,\overline{z_5})(z_5,\overline{z_1}),\non \\ \eea
two of which are depicted by figure 4. 
We have refrained from drawing $\mathcal{B}_{13}$, as it looks exactly the same as the graph $\mathcal{B}_9$ depicted by figure 3. However, these two graphs are different. This can be easily seen by assigning arrows to the dressing factors in the graphs (which we have ignored for the sake of brevity) such that in $(w,\overline{z})$ the arrow goes from the vertex $w$ to the vertex $z$ along the dashed line, which gives an orientation to each closed loop formed by the dressing factors. Then we see that the orientations of the two loops formed by the dressing factors in the graphs $\mathcal{B}_9$ and $\mathcal{B}_{13}$ are different. In fact, all the other graphs that arise in our analysis do not have this ambiguity and are uniquely defined without the need to specify the orientation of the loops.

To proceed, we rewrite $\mathcal{B}_{12}$ as
\bea \label{def12}\mathcal{B}_{12} &=&  \int_{\S^6} \prod_{i=1}^6 d^2 z_i G(z_1,z_2)G(z_3,z_4)G(z_5,z_6)(z_1,\overline{z_6})(z_6,\overline{z_2})\non \\ &&\times \frac{\Delta(z_2,z_3)\Delta(z_4,z_5)\overline{\Delta(z_3,z_1)}\overline{\Delta(z_5,z_4)}}{({\rm det}Y)^2}. \eea
Using \C{relDel} for $\Delta(z_4,z_5)\overline{\Delta(z_5,z_4)}/{\rm det}Y$ and $\Delta(z_2,z_3)\overline{\Delta(z_3,z_1)}/{\rm det}Y$, we obtain the expression given in \C{defBmore}. We next evaluate \C{def12} using \C{relDel} for $\Delta(z_4,z_5)\overline{\Delta(z_3,z_1)}/{\rm det}Y$ and $\Delta(z_2,z_3)\overline{\Delta(z_5,z_4)}/{\rm det}Y$ instead. Equating the resulting expression with $\mathcal{B}_{12}$ in \C{defBmore} gives the relation
\be \mathcal{B}_6 = \frac{1}{2}\mathcal{B}_1 \mathcal{B}_3\ee
between the graphs.

We next consider $\mathcal{B}_{13}$ which we rewrite as
\bea  \label{def13}\mathcal{B}_{13} &=&  \int_{\S^6} \prod_{i=1}^6 d^2 z_i G(z_1,z_2)G(z_3,z_4)G(z_5,z_6)(z_5,\overline{z_1})(z_2,\overline{z_6})\non \\ &&\times \frac{\Delta(z_1,z_3)\Delta(z_6,z_4)\overline{\Delta(z_3,z_5)}\overline{\Delta(z_4,z_2)}}{({\rm det}Y)^2}.\eea
Using \C{relDel} for $\Delta(z_1,z_3)\overline{\Delta(z_3,z_5)}/{\rm det}Y$ and $\Delta(z_6,z_4)\overline{\Delta(z_4,z_2)}/{\rm det}Y$, this reduces to the expression in \C{defBmore}.
We now evaluate \C{def13} differently using \C{relDel} for $\Delta(z_1,z_3)\overline{\Delta(z_4,z_2)}/{\rm det}Y$ and $\Delta(z_6,z_4)\overline{\Delta(z_3,z_5)}/{\rm det}Y$. Equating the resulting expression with $\mathcal{B}_{13}$ in \C{defBmore} yields
\be \label{7}\mathcal{B}_7 = \frac{1}{2}\mathcal{B}_1\mathcal{B}_4.\ee
Now the relations \C{rel711}, \C{rel78}, \C{7} immediately gives us that
\be \label{8}\mathcal{B}_8 =0.\ee

In fact let us deduce the relation \C{8} directly. To see this rewrite $\mathcal{B}_{14}$ in \C{defBmore} as
\bea  \label{def14}\mathcal{B}_{14} &=&  \int_{\S^6} \prod_{i=1}^6 d^2 z_i G(z_1,z_2)G(z_3,z_4)G(z_5,z_6)(z_1,\overline{z_3})(z_3,\overline{z_2})\non \\ &&\times \frac{\Delta(z_2,z_6)\Delta(z_4,z_5)\overline{\Delta(z_6,z_4)}\overline{\Delta(z_5,z_1)}}{({\rm det}Y)^2}.\eea
Using \C{relDel} for $\Delta(z_2,z_6)\overline{\Delta(z_6,z_4)}/{\rm det}Y$ and $\Delta(z_4,z_5)\overline{\Delta(z_5,z_1)}/{\rm det}Y$, we see it reduces to the expression in \C{defBmore}. Evaluating \C{def14} using \C{relDel} for $\Delta(z_2,z_6)\overline{\Delta(z_5,z_1)}/{\rm det}Y$ and $\Delta(z_4,z_5)\overline{\Delta(z_6,z_4)}/{\rm det}Y$ instead, and equating the resulting expression with $\mathcal{B}_{14}$ in \C{defBmore} gives back \C{8}.

\providecommand{\href}[2]{#2}\begingroup\raggedright\endgroup


\end{document}